\documentclass[preprint,12pt]{elsarticle}
\usepackage{setspace}
\usepackage{amssymb}
\usepackage{amsmath}

\usepackage{icarus}

\usepackage{natbib}

\bibpunct{(}{)}{;}{a}{,}{,}

\usepackage{graphicx}
\newcommand{\de}{$^{\circ}$}
\newcommand{\methane}{CH$_4$}
\newcommand{\ms}{m~s$^{-1}$}
\usepackage{rotating}

\begin{document}
\doublespacing

\begin{frontmatter}



\title{A 3D general circulation model for Pluto and Triton with fixed volatile abundance and simplified surface forcing}


\author[zalucha]{Angela M. Zalucha} 
\author[zalucha]{Timothy I. Michaels}

\address[zalucha]{SETI Institute, 189 Bernardo Ave., Suite 100, Mountain View, CA, 94043, USA}

%% This copyright statement isn't required at any stage by the Icarus
%% Editorial Office or Elsevier.  However, until you sign over the
%% copyright to Elsevier prior to publication (or negotiate with them
%% about copyright), you own the copyright to anything you create.
%% Just to keep things unambiguous, always include a copyright statement
%% or explicitly dedicate your work to the public domain.
%this for some reason is displaying on the front page, not under the author list.  Since
%it's not required, comment it out. -amz
%\begin{center}
%\scriptsize
%Copyright \copyright\ 2005, 2006 Ross A. Beyer, David P. O'Brien, Paul
%Withers, and Gwen Bart
%\end{center}

%% ----- ELSEVIER STUFF -----
%% The commands below up to the \end{frontmatter} are commented out
%% so that we can do some Icarus-required formatting on the second and
%% third pages that is not required later on by Elsevier.  So when
%% your paper gets accepted, and you are ready to start dealing with
%% Elsevier, copy your abstract and keywords up here, uncomment these
%% lines, and comment out the ICARUS STUFF below.
%% 
%% Alternately, you might just want to move these abstract, keyword,
%% and end frontmatter commands down, and comment out the ICARUS STUFF
%% commands.  It doesn't matter.

%\begin{abstract}
% % Text of abstract
% 
% \end{abstract}
% 
% \begin{keyword}
% % keywords here, in the form: keyword \sep keyword
% 
% 
% % PACS codes here, in the form: \PACS code \sep code
% 
% \end{keyword}

%% ----- END ELSEVIER STUFF -----

\end{frontmatter}

%% ----- ICARUS STUFF -----
%% Some formatting on the first, second, and third pages are required
%% by the Icarus Editorial Office that are not required by Elsevier.
%% This section contains those things.  When you are ready to transition
%% to ``Elsevier'' mode, copy your abstract and keywords up into
%% the ELSEVIER STUFF section, and then you can just delete everything
%% in this section.

%% We need to list the number of manuscript pages, figures, and tables. 
%%
%% Rather than manually count these things out, we'll use a little
%% trick here from Paul.  All you have to do is place three \label{}
%% tags on the last page, the last table, and the last figure, that
%% way these values are automatically updated (as long as you remember
%% to move the lasttable and lastfig labels when you add or remove
%% tables and figures).

\begin{flushleft}
\vspace{11.0 cm} %adjust this manually so that the following text appears at the bottom of page 1. -amz
%Number of pages: \pageref{lastpage} \\
% have no idea why it can't find lastpage label.  Now I have to hardwire it.
Number of pages: 58 \\
%Number of tables: \ref{lasttable}\\
Number of tables: 5 \\
%Number of figures: \ref{lastfig}\\
Number of figures: 13
\end{flushleft}

%% Don't worry about finding the various last* tags and deleting them
%% when you go to ``Elsevier'' mode if you don't want to, they should be
%% silently ignored.

%% The second page should indicate a proposed running head of not more 
%% than 55 characters, and the name and address to which editorial 
%% correspondence and proofs should be directed.  The pagetwo 
%% environment that icarus.sty provides will make page two for you,
%% just give the running head as an argument to the environment, and
%% then your correspondence address inside.
\begin{pagetwo}{3D Pluto and Triton general circulation model}
%                        1         2         3         4         5
%               1234567890123456789012345678901234567890123456789012345

\noindent Angela M. Zalucha\\
1050 Walnut Street\\
Suite 300\\
Boulder, CO 80302, USA \\
\\
Email: azalucha@seti.org\\
Phone: (720) 208-7211
%Fax: (650) 604-6779

\end{pagetwo}

\begin{abstract}
We present a 3D general circulation model of Pluto and Triton's atmospheres, which uses radiative-conductive-convective forcing.  In both the Pluto and Triton models, an easterly (prograde) jet is present at the equator with a maximum magnitude of 10--12~\ms and 4~\ms, respectively.  Neither atmosphere shows any significant overturning circulation in the meridional and vertical directions.  Rather, it is horizontal motions (mean circulation and transient waves) that transport heat meridionally at a magnitude of 1 and 3~$\times 10^7$ W at Pluto's autumn equinox and winter solstice, respectively (seasons referenced to the Northern Hemisphere).  The meridional and dayside-nightside temperature contrast is small ($\le$5~K).  We find that the lack of vertical motion can be explained on Pluto by the strong temperature inversion in the lower atmosphere.  The height of the Voyager 2 plumes on Triton can be explained by the dynamical properties of the lower atmosphere alone (i.e., strong wind shear) and does not require a thermally defined troposphere (i.e., temperature decreasing with height at the surface underlying a region of temperature increasing with height).  The model results are compared with Pluto stellar occultation light curve data from 1988, 2002, 2006, and 2007 and Triton light curve data from 1997.

\end{abstract}

% %% Keywords should appear after the abstract. 

%% ----- END ICARUS STUFF -----

%main text

\section{Introduction}
Pluto and Triton are sometimes referred to as ``sister'' worlds.  They have similar radii of 1132--1200~km for Pluto (Young and Binzel, 1994; Tholen and Buie, 1990, 1997; Buratti et al., 1995; Reinsch and Festou, 1994; Zalucha et al. 2011a,b) and 1353~km for Triton~\citep{thomas:2000}.  Their atmospheric compositions are primarily N$_2$ with trace amounts of \methane~and CO (and CO$_2$ for Triton)~\citep{owen:1993,young:1997,greaves:2011,herbert:1991,cruikshank:1993,lellouch:2009,lellouch:2010,lellouch:2011}.  Both have a rotation rate of about 6 Earth days and mean surface pressure of tens of microbars (Broadfoot et al., 1989; Lellouch et al., 2009; Zalucha and Gulbis, 2012).  Additionally, both receive weak insolation due to their large distance from the Sun.  Thus, their atmospheres are in very similar regimes (unlike Mars and Venus, or Earth and Titan), and we may apply the same model framework to both of these bodies.

A large amount of analysis and modeling has been done on the 1D vertical thermal structure of Pluto and Triton's atmospheres and surfaces
(e.g., Hansen and Paige, 1992,1996; Yelle and Lunine, 1989; Hubbard et al., 1990; Stansberry et al., 1994; Strobel et al., 1996; Elliot and Young, 1992; Elliot et al., 2003; Zalucha et al., 2011a,b).  While it is mathematically possible that these atmospheres' vertical temperature structure is set by energetic processes alone, it is rather unlikely that Pluto and Triton have no significant atmospheric circulation based on observations and physical theories of planetary atmospheres.   Though somewhat inhibited by the relative difficulty in observing atmospheric circulation, little work has been done exploring the atmospheric circulation of Pluto and Triton.  Occultation studies of Pluto have provided some measure of the wind speed~\citep{person:2008} and gravity waves~\citep{hubbard:2009} on Pluto, while the Voyager 2 flyby of Triton resulted in observations of surface-based vertical jets of aerosols~\citep{smith:1989} that were sheared off aloft by high altitude winds~\citep{soderblom:1990}.  Subsequent analysis was performed on the Voyager 2 observations of surface wind streaks by \citet{hansen:1990} and plume dynamics by \citet{ingersoll:1990}.

Recently, numerically sophisticated general circulation models (GCMs) have been applied to Pluto (Zalucha and Gulbis, 2012; Michaels and Young, 2011) and Triton~\citep{mueller-wodarg:2001,vangvichith:2011,miller:2011}.  GCMs not only include the physics of fluid motion, but are also coupled with surface processes and thermodynamical processes such as radiation, conduction, and condensation.  Despite the sudden proliferation of Pluto GCMs (PGCMs) and Triton GCMs (TrGCMs), all currently existing ones use different physical assumptions and a consensus about Pluto's atmospheric circulation is far from being reached.  Some of the difficulty may be attributed to Pluto and Triton's steep, low-altitude temperature inversion~\citep{yelle:1989,hubbard:1990,strobel:1996}, which makes analogies between other Solar System bodies such as Mars and Titan difficult beyond the very simplest arguments.

This paper continues the work of Zalucha and Gulbis (2012, hereinafter referred to as ZG12) from a 2D PGCM model to a 3D PGCM and also applies this particular model to Triton for the first time.  The ZG12 PGCM/TrGCM uses the dynamical core of the Massachusetts Institute of Technology (MIT) GCM~\citep{marshall:1997} and radiative-conductive forcing scheme of \citet{yelle:1989} (now augmented with the convective forcing of Zalucha et al., 2011b).  The PGCM/TrGCM makes implicit assumptions about the surface, subsurface, and condensation cycle (elaborated in ZG12).  Surface and subsurface parameters (such as surface emissivity, surface albedo, surface thermal inertia, and surface frost history) are not well constrained and represent a large area of parameter space that is not possible to analyze in a single paper and is left for future work.

Stellar occultation observations represent a significant fraction of the available data for Pluto, and for Triton to a lesser extent.  We use the forward method for validating the PGCM and TrGCM with the data; namely, we take pressure-temperature profiles from the PGCM/TrGCM, calculate a model light curve, and assess any differences in normalized flux vs. radius (or time) space.  Temperature profiles derived from inverse methods \citep{elliot:2003b} may contain systematic errors due to the assumption of an upper boundary condition, while idealized methods \citep{elliot:1992} must make assumptions about the variation of temperature with height.  The GCM to light curve conversion has been used numerous times (Zalucha et al., 2011a,b; ZG12) and we are increasingly confident of this technique.

In Section~\ref{se:model} we describe our PGCM and TrGCM setup and briefly summarize the light curve model.  In Section~\ref{se:results}, we compare the PGCM and TrGCM output with stellar occultation light curves and present the corresponding temperature and zonal wind fields.  We also show the meridional heat transport on Pluto.  In Section~\ref{se:discussion} we discuss the effect of Pluto's lapse rate on atmospheric circulation and provide evidence that Triton's lower atmospheric structure is dynamically (rather than thermally) maintained.
	
\section{Model setup}\label{se:model}
	\subsection{General circulation model}
Both the PGCM and the TrGCM use the same dynamical core, based on the MIT GCM.  The configuration of the dynamical core was described in ZG12 and will only be briefly summarized here.  The MIT GCM solves the primitive equations of geophysical fluid dynamics (i.e., the Navier-Stokes equations) under the assumptions of conservation of mass, the ideal gas law, and conservation of energy in the presence of an external source and sink term) on a sphere using the finite volume method.  The model is hydrostatic in the vertical and compressible.  The vertical grid is based on an $\eta$ coordinate and the surface topography is flat.  Boundary layer friction is represented by a simple drag law (linearly dependent on the horizontal velocity) that decreases with height, reaches zero at the top of the boundary layer, and is zero at all levels above.  The composition of the atmosphere is assumed to be primarily N$_2$, with trace amounts of \methane~that is radiatively active but not advected.  In this version of the model, mass is not permitted to be exchanged between the surface and the atmosphere (but will be in future versions).  When the atmospheric temperature is diagnosed to fall below the N$_2$ freezing temperature, it is instantaneously reset back to the freezing temperature.  Surface thermal inertia and albedo are globally and temporally constant (this assumption will also be relaxed in the future).  Surface temperature is held at the N$_2$ freezing temperature, consistent with observations that Pluto and Triton are frost-covered~\citep{owen:1993,tryka:1993}.  The implications of this assumption are discussed in Section~\ref{ss:lcresults}.  The latitude and longitude coordinates we use to describe locations on Pluto and Triton's surfaces follow the convention that Pluto's north pole is in the same hemisphere as the ecliptic north pole with longitude is increasing eastward.  This convention is equivalent to stating that both Pluto and Triton rotate in a retrograde fashion (i.e., clockwise when viewed from north).

An important improvement since ZG12 is the expansion of the model domain to three spatial dimensions.  In the horizontal, the MIT GCM uses a cubed-sphere grid~\citep{adcroft:2004} with 32 $\times$ 32 points per cube face, equivalent to a grid spacing of 2.8$^{\circ}$ at the equator.  Compared to the more common cylindrical projection grid (i.e., a latitude/longitude grid), this type of horizontal grid eliminates singularities at the poles that force meridional winds to zero and removes the requirement for Fourier filtering in the high latitudes (in order to maintain a practical timestep). 

As in ZG12, we use the radiative-conductive scheme of \citet{yelle:1989} to calculate the external heating and cooling terms due to \methane~and molecular conduction (now including the longitudinal dependence in insolation).  This scheme captures non-local thermodynamic equilibrium (non-LTE) effects and the sharp stratospheric temperature inversion, while being computationally practical in a GCM.  Voyager 2 observations indicate the existence of a troposphere on Triton \citep{smith:1989}, so we have modified the \citet{yelle:1989} model to include this effect following the procedure described in Zalucha et al. (2011b).  The heat balance equation is
\begin{equation}
\rho c_p\frac{\partial T}{\partial t} = Q - l +\frac{\partial}{\partial z}\left(k_o T^{\alpha}\frac{\partial T}{\partial z}\right) +C_{diff}\label{eq:heatbalance},
\end{equation}
	where $\rho$ is atmosperic density, $c_p$ is the specific heat at constant pressure, $T$ is temperature, $z$ is altitude, the third term on the RHS is the conduction term with thermal conductivity coefficient $k_o$ and thermal conductivity exponent $\alpha$, $Q$ is the non-LTE heating by CH$_4$ at 3.3~$\mu$m, and $l$ is the non-LTE cooling by CH$_4$ at 7.8~$\mu$m.  The \citet{yelle:1989} model used to specify $Q$ and $l$ does not include CO, which has been recently observed in Pluto's atmosphere~\citep{greaves:2011,lellouch:2011}.  It is difficult to analytically express the cooling effects of CO; thus CO has not been included at this stage owing to computational simplicity.  We plan to add the effects of CO using the \citet{strobel:1996} radiative-conductive model in future work.	
	
	In Eq.~\ref{eq:heatbalance}, $C_{diff}$ is the term that prescribes the troposphere via convection or
	\begin{equation}
	C_{diff}=\left(\frac{\partial \theta_e}{\partial T}\right)^{-1}\frac{\partial}{\partial z}\left(K_c\frac{\partial \theta_e}{\partial z}\right),
	\end{equation}
	where $\theta_e$ is the equivalent potential temperature (due to condensation or sublimation of N$_2$) and $K_C$ is the convection diffusion coefficient.  Note that here we have used equivalent potential temperature assuming implicit condensation and sublimation of N$_2$ in order to mathematically represent the troposphere.  As stated previously, no explicit exchange of mass occurs in the continuity equation (in the model's dynamical core).  Assuming that the tropospheric temperature is primarily determined by convection, the vertical temperature profile will be equal to the moist (with respect to N$_2$) adiabatic lapse rate of $-dT/dz=0.09$~K~km$^{-1}$.  Specifically,
	\begin{equation}\label{eq:kcfunc}
K_C=20\left[1-\tanh\left(\frac{z-h_c}{5}\right)\right],
\end{equation}
where $z$ has units of km and $h_c$ is the troposphere critical height (in km), interpreted as the altitude to which convective effects are significant.  $h_c$ varies monotonically with the depth of the troposphere (i.e., larger $h_c$ corresponds to larger troposphere depth), but the two are not necessarily equal.  The functional form of Eq.~\ref{eq:kcfunc} was chosen based on the structure of Earth's troposphere as in Zalucha et al. (2011b).

It has been known for many decades that terrestrial climate models need a momentum sink in the mesosphere to match observations.  Initially, this was parameterized with Rayleigh friction~\citep[e.g.,][]{leovy:1964}, a momentum sink that was linearly dependent on velocity and increased in magnitude with height.  More elaborate schemes varied this forcing in time and space until the desired result was achieved.  Eventually, it was discovered that the breaking of subgrid-scale gravity waves (also known as buoyancy or internal waves) was the physical mechanism responsible for the momentum drag~\citep{houghton:1978,lindzen:1981,holton:1982,holton:1983,matsuno:1982}.  While the theory behind linear, small-amplitude, idealized gravity waves is well-understood, these waves present a problem for global climate models because their typical wavelength is smaller than the grid spacing.  Hence, the amount of momentum deposition due to breaking waves must be parameterized.  

Various gravity wave parameterizations have been implemented for Earth \citep[e.g.,][]{lindzen:1968,lindzen:1981}, Mars~\citep[e.g.,][]{medvedev:2011}, and Venus (e.g., Zalucha et al., 2013) with varying degrees of success.  In all cases, the parameterizations must be tuned to observations (usually of perturbations in vertical profiles of temperature or density).  Observations of such waves on Pluto and Triton are almost nonexistent~\citep[but see][for one example]{hubbard:2009}.  Thus, in the PGCM and TrGCM we must resort to Rayleigh friction.  For numerical stability, a surprisingly large momentum sink is required in the uppermost five levels of the model domain.  In order of decreasing height, the drag coefficients are 0.02083~s$^{-1}$, 0.00694~s$^{-1}$, 0.00231~s$^{-1}$, 0.00072~s$^{-1}$, and 0.00024~s$^{-1}$ ; for comparison, typical values for Mars GCMs in the uppermost three levels are $10.1\times10^{-5}$~s$^{-1}$, $3.4\times10^{-5}$~s$^{-1}$, and $1.1\times10^{-5}$~s$^{-1}$.  The 2D PGCM did not require such strong friction (in fact it was stable even with no Rayleigh friction).  The reason for such strong Rayleigh friction might indicate that vertically propagating gravity waves deposit relatively large amounts of momentum in Pluto and Triton's upper atmosphere.  Were we able to incorporate a true gravity wave parametrization (i.e., if we had observational information about the wave structure on Pluton and Triton), the strength of the drag would potentially depend on longitude and latitude.

In practice, the Rayleigh friction layer (often referred to as the sponge layer) is not treated as giving physically meaningful results, and the model output in the sponge layer is ignored.  Likewise, we do not present results from this layer in our results (Section~\ref{se:results}).  The effect of choosing the top five layers as the sponge layer, rather than the top four or six, simply determines the altitude to which we are able to ascertain the circulations of Pluto and Triton's atmosphere.  In the 2D version of the PGCM, a thicker sponge layer led to weaker zonal winds.  We have not performed a detailed analysis on the effect of the strength of the momentum sink in either the 2D or the 3D PGCM.  As stated above, the 2D PGCM was stable without a sponge layer at all, which is due to the fact that zonally propagating waves are not possible in a 2D GCM.

Both the PGCM and TrGCM were run for a 15 Earth-year spin up period (as explained in ZG12), then continued through the year 2007.  Since the spin up time is determined primarily by the radiative-conductive relaxation rate, the addition of a third spatial dimension does not change the spin up time.  The first occultation dataset for Pluto was obtained in the year 1988~\citep{elliot:1989}, so the PGCM simulations begin in the year 1973.  Fewer occultation datasets exist for Triton; the highest signal-to-noise ratio (SNR) dataset was obtained in 1997~\citep{elliot:1998}, so the TrGCM simulations begin in the year 1982.   In both cases, the atmospheres were initialized in radiative-conductive-convective equilibrium (i.e., steady state solution to the radiative-conductive-convective scheme with no wind) and with globally constant surface pressure.

Table~\ref{tb:params} shows the bulk parameters of the PGCM and TrGCM.  For Pluto, we performed a parameter sweep of surface pressure in the range of 4--28~$\mu$bar in intervals of 4~$\mu$bar.  The \methane~mixing ratio was set at 0.2$\%$, 0.6$\%$, and $1\%$.  The ranges and intervals were chosen in accordence with previous analyses and observations~\citep[e.g., ZG12,][]{lellouch:2009}.  We determined the best-fit surface pressure/\methane~mixing ratio parameter combinations for a particular observational dataset by calculating a model stellar occultation light curve from each PGCM simulation (see Section~\ref{ss:lcm}) and then finding the model run that has the minimum reduced $\chi^2$ compared to the data.  The existence of a troposphere on Pluto is debated.  \citet{stansberry:1994} showed that the addition of a troposphere caused a kink in the light curve \citep[similar to that observed in the 1988 occultation by][]{elliot:1989}.  The flux level of the kink and depth of the light curve could be changed by varying the depth and lapse rate of the troposphere.  \citet{stansberry:1994} show that a troposphere was consistent with the data, but not required.  Moreover, Zalucha et al. (2011b) find better agreement for a model with no troposphere.  Thus, we assume no troposphere in the PGCM.

Triton was visited by Voyager 2 in 1989, so we used measured values from this encounter, along with more recent measurements, to configure the TrGCM.  We then compare the results to the 4 November 1997 stellar occultation observed by the Hubble Space Telescope (HST) \citep{elliot:1998}, which is the only Triton occultation dataset with the minimum SNR threshold to be used in our analysis (see Zalucha et al., 2011b for a definition of the minimum SNR).

During the Voyager 2 encounter with Triton in 1989 the atmospheric pressure was determined to be 14~$\mu$bar~\citep{broadfoot:1989}. The \methane~partial pressure was determined to be 2.45~nbar~\citep{herbert:1991,strobel:1995}.  \citet{elliot:1998} observed from a stellar occultation that occurred on 4 November 1997 that the pressure at the half light radius had increased significantly since the Voyager 2 encounter and measurements from a 1995 stellar occultation.  They concluded that the surface pressure of Triton doubles every 10 years during the 1989--1997 period of observation.  Using spectroscopic measurements, \citet{lellouch:2010} found  that in 2009 the partial pressure of \methane~had increased to $9.8 \pm 3.7$~nbars.  Combining all this information and performing a linear interpolation, we infer that the surface pressure in 1997 was 25.2~$\mu$bar and the mixing ratio of \methane~was 0.02$\%$.  Voyager 2 also observed plumes rising from the surface and then being sheared off at an altitude of approximately 8~km~\citet{smith:1989}.  This has been interpreted as the depth of a thermally-defined troposphere~\citep{yelle:1995}.  Since the tropospheric depth has not been observed since the Voyager 2 encounter, we take the troposphere depth in the TrGCM to be 8~km.  We reiterate that our method is not simply a reanalysis of Voyager 2 derived results, but synthesizes it along with more recent measurements~\citep{elliot:1998,lellouch:2010}.

	\subsection{Light curve model}\label{ss:lcm}
	The PGCM and TrGCM output pressure-temperature vertical profiles as a function of latitude, longitude, and time.  Using hydrostatic balance, the ideal gas law, and the linear dependence between number density and refractivity, the refractivity $\nu$ as a function of the body's radius $r$ may be obtained.  Then the bending angle $\theta$ and its vertical derivative $d\theta/dr$ is given by \citet{chamberlain:1997}:
	\begin{equation}
	\theta(r)=\int^{\infty}_{-\infty}\frac{r}{r'}\frac{d\nu(r')}{dr'}dx
	\end{equation}
	and
	\begin{equation}
	\frac{d\theta(r)}{dr}=\int^{\infty}_{-\infty}\left[\frac{x^2}{(r')^3}\frac{d\nu(r')}{dr'}+\frac{r^2}{(r')^2}\frac{d^2\nu(r')}{dr'^2}\right]dx,
	\end{equation}
	where $x^2=r'^2-r^2$ is the path along the ray and $r'$ is a dummy variable of integration.  The normalized light curve flux is given by (assuming no line of sight extinction)
	\begin{equation}\label{eq:lcflux}
\zeta = \frac{1}{\left\vert 1+D[d\theta(r)/dr]\right\vert }\frac{1}{\left\vert 1+D\theta(r)/r\right\vert},
\end{equation}
where $D$ is the distance between the observer and body.  In this model, we have assumed no line of sight extinction along the incoming stellar light ray.  See Zalucha et al. (2011a) for a further explanation of this method.

The forward method of light curve modeling is employed, where the GCM results are used as input into the \citet{chamberlain:1997} scheme, yielding a model, GCM-based light curve.  The agreement between data and model is then evaluated in flux vs. radius (or flux vs. time) space.  Idealized light curve modeling techniques provide temperature as a function of radius from light curve data.  However, these techniques \citep[such as][]{elliot:1992} must make an assumption about the temperature structure (in this case that is follows a power law), which is usually extremely simplified for mathematical convenience and thus not accurate for planetary atmospheres in general.  Other techniques such as the inversion method~\citep{elliot:2003b} require an upper boundary condition, which is usually obtained from an idealized model and can introduce large systematic errors.  Forward modeling can utilize atmospheric fields having any vertical structure and is therefore potentially more accurate.

	\section{Results}\label{se:results}
	\subsection{Best-fit PGCM light curves}\label{ss:lcresults}
	We validate our PGCM using the following datasets, which are selected based on the criterion that they have the highest SNR per scale height (see Zalucha et al., 2011a):  9 June 1988 on the Kuiper Airborne Observatory (KAO)~\citep{elliot:1989} 0.9~m telescope; 21 August 2002 at Mauna Kea on the University of Hawaii 2.2~m telescope (UH 2.2m)~\citep{elliot:2003a,pasachoff:2005}; 12 June 2006 at Siding Spring, Australia on the 2.3~m Australian National University (ANU) telescope~\citep{elliot:2007} and on the 4~m Anglo-Australian Telescope (AAT)~\citep{eyoung:2008}; and 31 June 2007 at Mt. John Observatory in New Zealand~\citep{olkin:2009}.
		
	In our current analysis	we also performed PGCM simulations in which the surface temperature was either free to evolve (i.e., determined by the insolation on an non-volatile surface) or fixed (i.e., held fixed at the N$_2$ freezing temperature).  The parameters of best-fit (i.e., minimum reduced $\chi^2$) are shown in Tables~\ref{tb:results_free} and \ref{tb:results_fixed}, respectively.  Note that these values assume a surface radius of 1152~km, discussed below.  The formal error bars (see ZG12 for details about this calculation) are smaller than the interval of the parameter sweep, so we take the interval of the parameter sweep to be the reported error bars.    The fixed surface temperature simulations always have a lower reduced $\chi^2$, indicating that they are better fits.  Moreover, the derived surface pressures for these runs agree better with previous analyses (Lellouch et al., 2009; Zalucha et al., 2011b).
	
	To put into perspective the differences in the reduced $\chi^2$ values of the fixed vs. free surface temperature simulations, we point out that in all years, the reduced $\chi^2$ value for adjacent simulations in parameter space (i.e., the one-sigma error bars) for the fixed surface temperature simulations was still less than the free surface temperature case.  For example, consider the reduced $\chi^2$ values for the 2007 fixed surface temperature case in the vicinity of the the best-fit values for surface pressure and \methane~mixing ratio (Table~\ref{tb:2007example}).  The highest reduced $\chi^2$ in this subset of parameter space is 2.386, which is lower than the 2007 free surface temperature minimum reduced $\chi^2$ value of 4.343.  Thus, at the very least we may claim that the free surface temperature results are worse than the one-sigma error bars on the fixed surface temperature results.  In this case, the disagreement is many sigma.
	
		In addition to the PGCM parameter sweep of global mean surface pressure and \methane~mixing ratio, we investigated three different surface radii spanning the bounds of published values: 1180~km (Zalucha et al., 2011a), 1152~km~\citep{elliot:2007}, and 1130~km~\citep{reinsch:1994}.  We found that in all cases the 1152~km light curves had the lowest minimum reduced $\chi^2$, which differs from the 2D PGCM where Z12 assumed a surface radius of 1180~km.  Table~\ref{tb:chi2vsrs} shows the $\chi^2$ values for these three radii.
	
	  The best-fit value for \methane~mixing ratio is 1$\%$ in all cases for fixed surface temperature, which agrees with ZG12 and~\citet{young:1997} but is larger than~\citet{lellouch:2009}.  The simulated surface pressure increases from 1988 to 2006 then drops slightly in 2007.  The trend agrees with \citet{elliot:2003a,elliot:2007}, Zalucha et al. (2011a), and ZG12. The values agree quantitatively with \citet{lellouch:2009}, Zalucha et al. (2011a), and ZG12.  We point out that our best-fit values for \methane~mixing ratio are at the upper boundary of \methane~mixing ratio values considered in this study.  However, the sensitivity of the radiative-conductive equilibrium temperature profile predicted by the \citet{yelle:1989} model decreases with increasing \methane~mixing ratio.  Above about 1$\%$, the model saturates, and little change is seen in the magnitude of the radiative-conductive equilibrium temperature above this value.  Thus, there is little to be gained by running the PGCM at \methane~mixing ratio values higher than 1$\%$.

\begin{figure}
\begin{center}
 \noindent\includegraphics[width=0.5\textwidth]{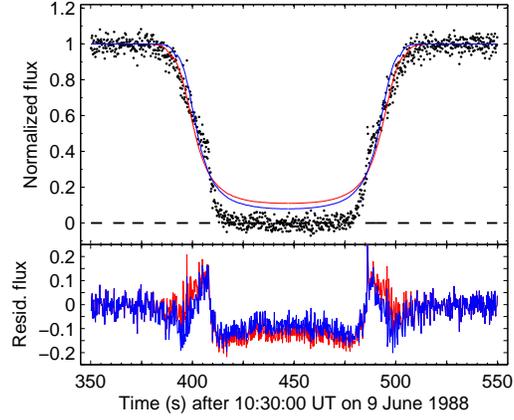}

 \caption{\label{fg:lc1988}  Top panel: Best PGCM fit for surface temperature free (red curve) and surface temperature fixed (blue curve) to 1988 KAO normalized light-curve data (black points).  The dashed horizontal line is the zero flux level.  Bottom panel: residual between models and data. }

 \end{center}
  \end{figure}

\begin{figure}
\begin{center}
 \noindent\includegraphics[width=0.5\textwidth]{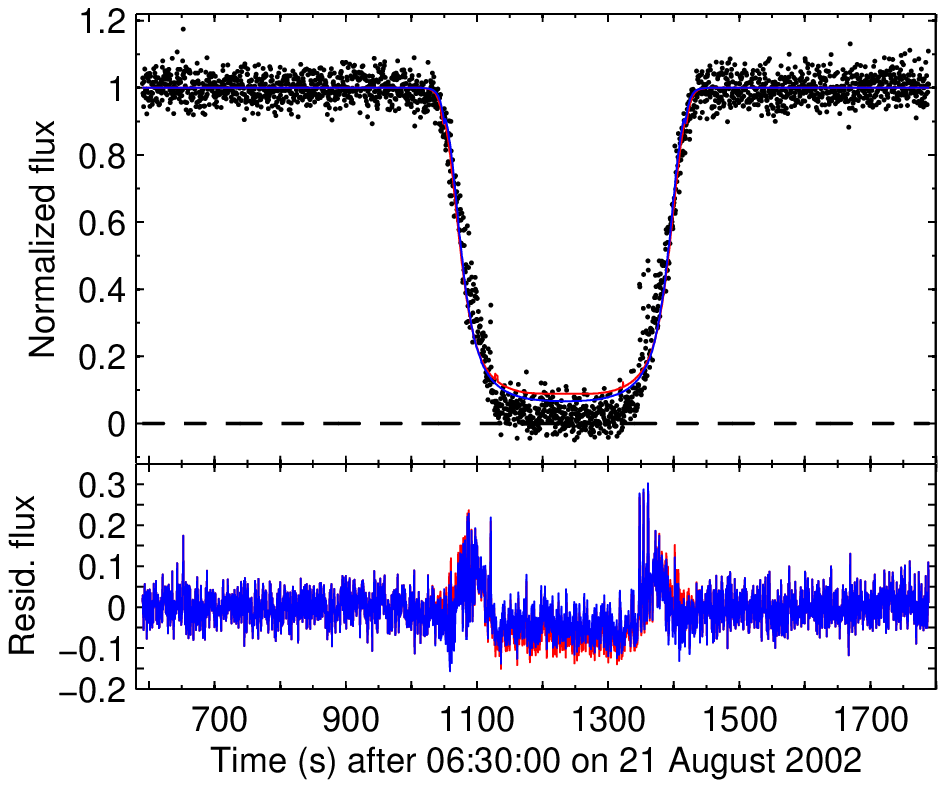}

 \caption{\label{fg:lc2002}  Top panel: Best PGCM fit for surface temperature free (red curve) and surface temperature fixed (blue curve) to 2002 UH 2.2m normalized light-curve data (black points).  The dashed horizontal line is the zero flux level.  Bottom panel: residual between data and models. }

 \end{center}
  \end{figure}

\begin{figure}
\begin{center}
 \noindent\includegraphics[width=0.5\textwidth]{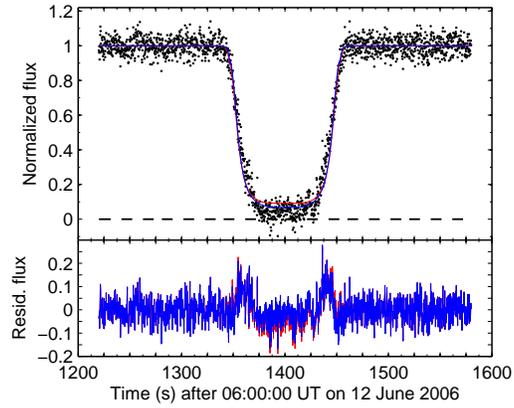}

 \caption{\label{fg:lc2006}  Top panel: Best PGCM fit for surface temperature free (red curve) and surface temperature fixed (blue curve) to 2006 Siding Spring normalized light-curve data (black points).  The horizontal vertical line is the zero flux level.  Bottom panel: residual between data and models. }

 \end{center}
  \end{figure}

\begin{figure}
\begin{center}
 \noindent\includegraphics[width=0.5\textwidth]{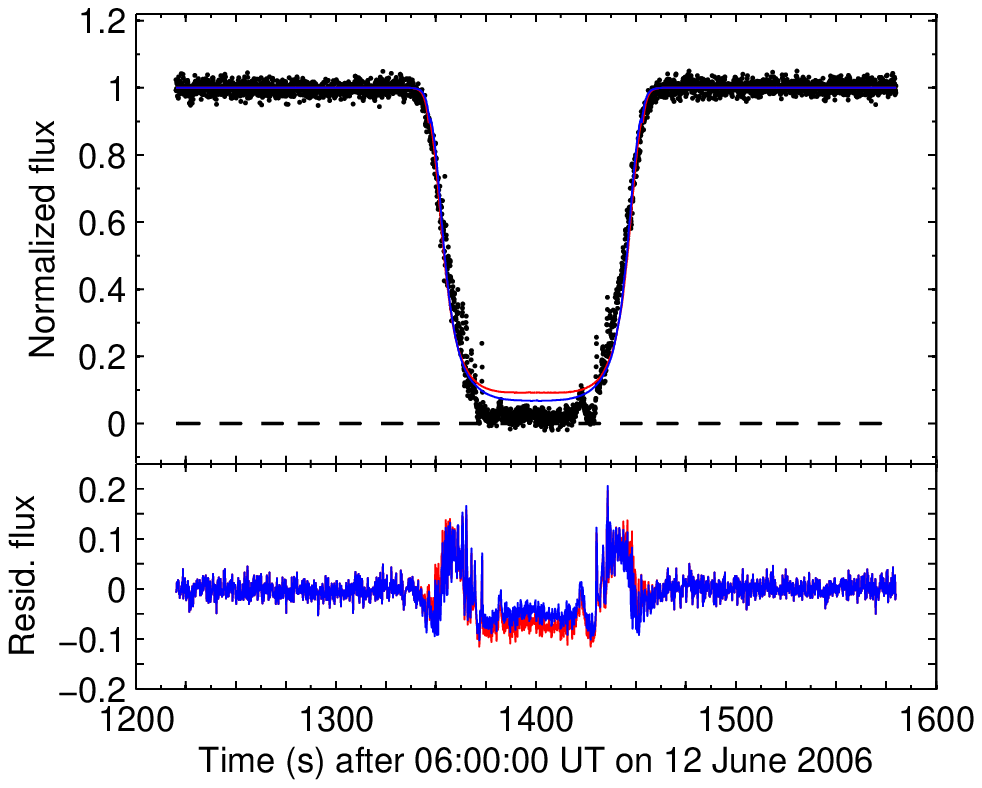}

 \caption{\label{fg:lc2006aat}  Top panel: Best PGCM fit for surface temperature free (red curve) and surface temperature fixed (blue curve) to 2006 AAT normalized light-curve data (black points).  The dashed horizontal line is the zero flux level.  Bottom panel: residual between data and models. }

 \end{center}
  \end{figure}

\begin{figure}
\begin{center}
 \noindent\includegraphics[width=0.5\textwidth]{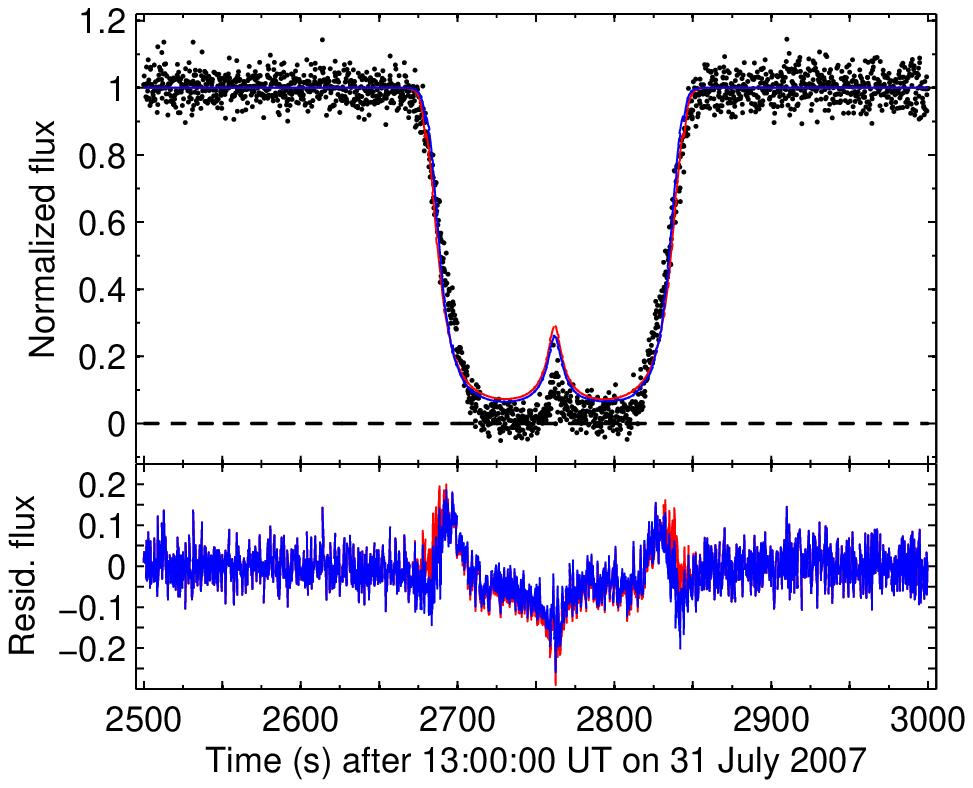}

 \caption{\label{fg:lc2007}  Top panel: Best PGCM fit for surface temperature free (red curve) and surface temperature fixed (blue curve) to 2007 Mt. John normalized light-curve data (black points).  The dashed horizontal line is the zero flux level.  Bottom panel: residual between data and models. }

 \end{center}
  \end{figure}

	Figures~\ref{fg:lc1988}--\ref{fg:lc2007} show the best-fit modeled and observed light curves.  The 3D PGCM light curves do not contain the oscillations that the 2D PGCM exhibited in ZG12, which were due to numerical artifacts in the 2D PGCM solution.  The models and data agree above the 0.4 normalized flux level.  At lower flux levels (i.e., closer to the center of the light curve), the model first underestimates the flux levels, then overestimates the flux around the midtime of the light curve.  The difference is approximately 10$\%$ or less for 2002, 2006, and 2007, when noise in the light curve is ignored.  The proximity of the mismatch to the midtime of the light curve indicates that Pluto's lower atmosphere is not being simulated accurately (the model light curves probe to a minimum altitude of about 20~km).  We may rule out the uncertainty in Pluto's surface radius as a cause of discrepancy, because \citet{zalucha:2011a} showed that the effect of changing light curve radius was to widen or narrow the model light curve; in fact the flux near the midtime of the light curve was affected very little by changing surface radius.  The assumption of fixed vs. free surface temperature affects the bottom of the light curve only indirectly through the assumption of energy balance within the column.  If the discrepancy were due to only haze extinction, then the model would overestimate the flux.  Since the model both overestimates and underestimates the flux near midtime, the model must instead be somehow misrepresenting the temperature gradient.  The model flux could be reduced near the midtime of the light curve if Pluto has a deep troposphere~\citep{stansberry:1994}, although it would contradict Zalucha et al. (2011b) who found no troposphere.  Note that the disagreement in 1988 is larger than the other years.  The cause of the sharp drop off (kink) at the 40$\%$ flux level has been well modeled by both a strong thermal inversion~\citep{lellouch:2009} and extinction~\citep{elliot:1992}, but to date no study has conclusively proven which scenario is taking place on Pluto.  
	
	\subsection{Best-fit PGCM wind, temperature, and surface pressure}
	 As in the 2D PGCM, the 3D PGCM solution contains essentially no vertical motions (the maximum magnitude of vertical velocity in pressure coordinates is of order 10$^{-7}$ $\mu$bar~s$^{-1}$).  Similarly, the surface pressure does not vary over the globe by more than 10$^{-3}$ $\mu$bar.  Unlike the 2D PGCM, the meridional motions are non-negligible, occurring in the form of transient waves.  The maximum magnitude of the instantaneous meridional velocity is 2~m~s$^{-1}$.
	 
\begin{figure}
\begin{center}
 \noindent\includegraphics[width=1.0\textwidth]{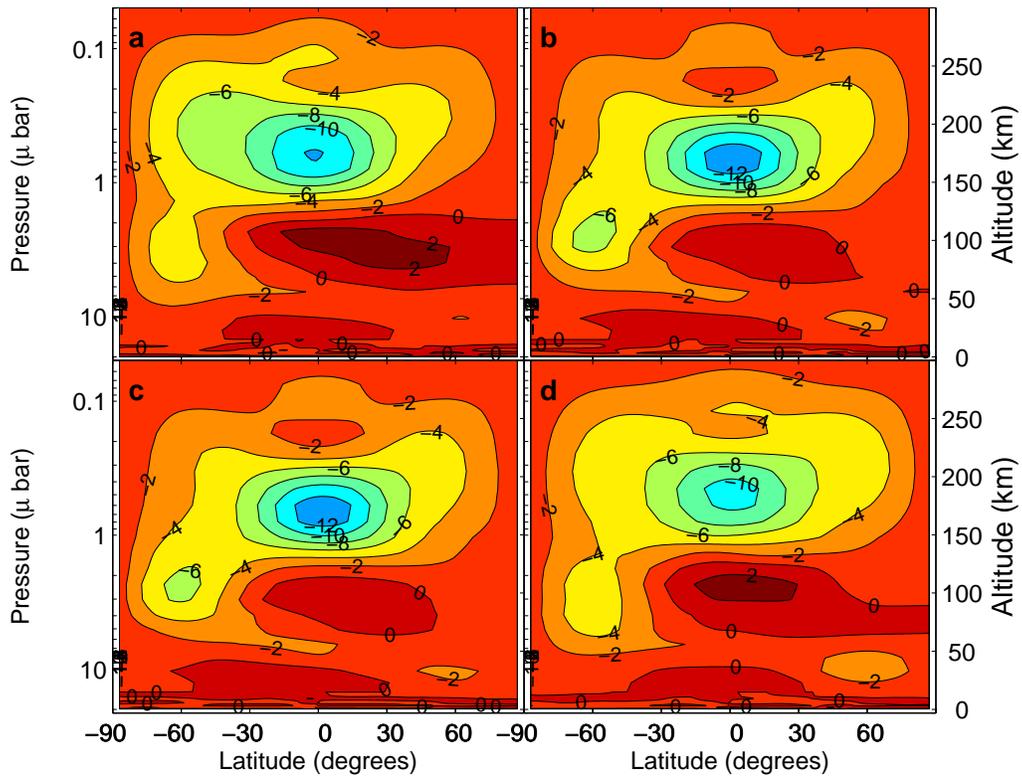}

 \caption{\label{fg:u}  Zonally averaged zonal wind (m~s$^{-1}$) corresponding to best-fit PGCM results from (a) 9 June 1988 (b) 21 August 2002 (c) 12 June 2006 (d) 31 July 2007.}

 \end{center}
  \end{figure}

	 The zonal wind is nearly homogeneous in longitude.  Figure~\ref{fg:u} shows the zonally averaged zonal wind corresponding to the dates of the occultation datasets.   Three vertical regimes exist in the modeled solution.  First is the near-surface (below 50~km altitude or above 10~$\mu$bar pressure) with weak winds except for a $\sim$2~\ms~jet centered at 60\de~latitude.  The winds in this region are weak because they lie at altitudes subject to surface frictional drag.  Above this region (between 50 and 130~km altitude and 2 and 10~$\mu$bar), the winds are easterly (maximum magnitude 6~\ms) between the South Pole and 30\de~latitude and westerly (maximum magnitude 2~\ms) at other latitudes.  Finally, in the region between 130~km and the base of the Rayleigh drag layer at 300~km (0.6 and 2~$\mu$bar), the winds are characterized by an easterly equatorial jet (maximum magnitude 12~\ms).  In the future we plan to perform a sensitivity study to determine if the depth of the model domain and/or vertical grid resolution affects the vertical boundaries of these regions and the associated wind magnitudes.  
	 
	 Few wind measurements exist for Pluto.  \citep{person:2008} derived an upper limit of 3~\ms, although as explained in ZG12, this analysis was not carried out correctly.  Moreover, it is unclear what altitude this value applies to, making direct comparison difficult.
	 
	The circulation pattern in the 3D PGCM is quite different than in the 2D PGCM, which exhibited a zonal circulation in cyclostrophic balance.  However, we would not expect the 2D and 3D circulations to be the same.  The 3D PGCM has a day-night insolation contrast which can drive significant circulations (in addition to the meridional contrast) and also allows for waves that can transport energy and momentum (and thus alter the circulation pattern)
	
\begin{figure}
\begin{center}
 \noindent\includegraphics[width=1.0\textwidth]{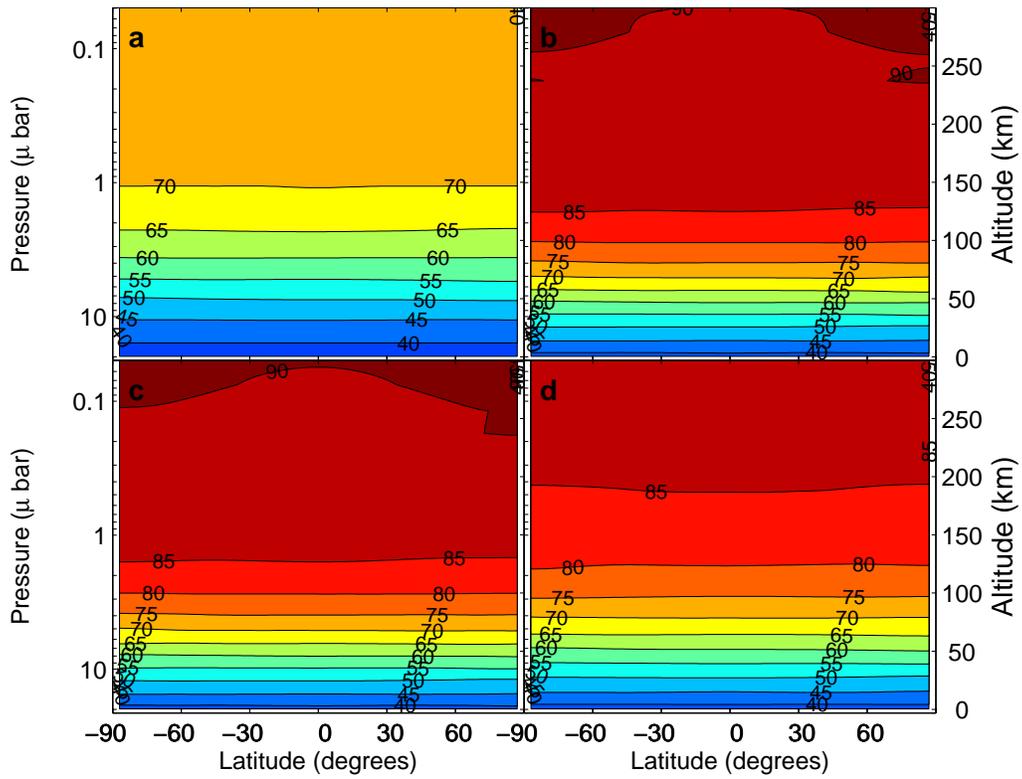}

 \caption{\label{fg:t}  Zonally averaged temperature (K) corresponding to best-fit PGCM results from (a) 9 June 1988 (b) 21 August 2002 (c) 12 June 2006 (d) 31 July 2007.}

 \end{center}
  \end{figure}

	Figure~\ref{fg:t} shows the zonally averaged temperature corresponding to the dates of the occultation datasets.  The global temperature field has few horizontal variations, despite the insolation varying in latitude and longitude.  The lack of horizontal variability in the temperature field suggests that the circulation is efficient at moderating the meridional and dayside-nightside heating gradients.  Pluto's rotation period is 6.4~Earth days, while the radiative relaxation time scale is of order 10 to 100 Earth days.  Thus the response of the atmosphere is much slower than the daily changes in radiative forcing, so we would not expect a strong day-night temperature contrast to exist.  \citet{yelle:1995}, using assumptions of a frost covered surface in vapor pressure equilibrium, predicted that the surface temperature and pressure would have very little variation on Triton (and Pluto by analogy).  This prediction is consistent with our GCM results, since the GCM is implicitly adding latent heat when the temperature drops below the N$_2$ freezing temperature and the temperature is reset back to the freezing temperature.  Future work will focus on the contribution of latent heating compared with other heat sources by modeling the volatile cycle explicitly.
		
	\subsection{Heat transport in Pluto's atmosphere}
	The meridional transport of energy $F_E$ is given by~\citep{peixoto:1992}
	\begin{equation}\label{eq:energy}
	\left[\overline{F_E}\right]=c_p\left[\overline{Tv}\right]+g\left[\overline{zv}\right]+\frac{1}{2}\left[\overline{\left(u^2+v^2\right)v}\right]+L\left[\overline{vq}\right],
	\end{equation}
	where brackets denote a zonal average, overbars denote a time average, $v$ is the meridional velocity, $g$ is the gravitational acceleration at the surface, $u$ is the zonal velocity, $L$ is the specific latent heat of sublimation, and $q$ is the specific humidity.  The terms on the right hand side of the equation, from left to right, are the transport of sensible heat, potential energy, kinetic energy, and latent heat.  Since we do not explicitly treat condensables, the latent heat term is zero.
	
\begin{figure}
\begin{center}
 \noindent\includegraphics[width=1.0\textwidth]{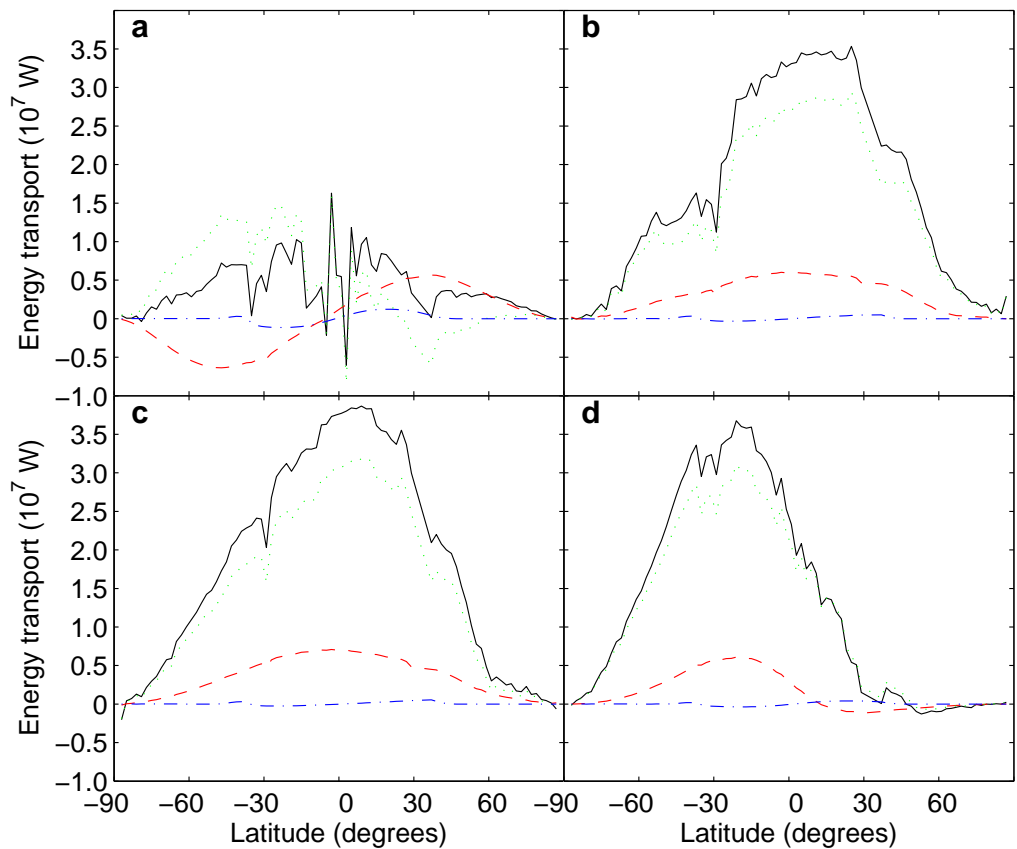}

 \caption{\label{fg:energy_physics}  Simulated vertically averaged meridional transport of energy on Pluto corresponding to (a) 9 June 1988 (b) 21 August 2002 (c) 12 June 2006 (d) 31 July 2007.  The black solid line is the total energy transport, the red dashed line is the potential energy transport, the green dotted line is the sensible heat transport, and the blue dashed-dotted line is the kinetic energy transport.  These terms are defined in Eq.~\ref{eq:energy} and have been multiplied by a factor of $2\pi a \cos\phi p_s$ (where $a$ is the surface radius, $\phi$ is latitude, and $p_s$ is surface pressure) to obtain units of Watts.  The results have been time averaged over 90 Earth days.  Positive values indicate northward transport of heat.}

 \end{center}
  \end{figure}

	Figure~\ref{fg:energy_physics} shows the PGCM values for vertically averaged meridional heat transport corresponding to the terms in Eq.~\ref{eq:energy}.  At the time of every occultation event, the kinetic energy is much smaller than the other terms, which is typical of other planetary atmospheres.  Pluto's equinox occurred in the year 1989, and its solstice will occur in 2030.  The equations for solar zenith angle as a function of latitude and season \citep[see e.g.,][]{peixoto:1992} show that bodies orbiting the Sun exhibit an equinoctial pattern for a comparatively short fraction of their orbit; the remaining time is spent in a solstitial forcing pattern.  Thus, by the year 2002, Pluto is already experiencing a solstitial forcing.  For the years 2002, 2006, and 2007, the energy transport is always positive.  In these years, the South Pole is the summer pole and the amount of diabatic (i.e., solar) heating decreases from south to north.  Thus, a positive energy transport corresponds to warm air moving northward, or a redistribution of energy from south to north.  This is consistent with the meridionally homogeneous temperature gradient seen in the PGCM temperatures (Fig.~\ref{fg:t}b--d).  Also, in the years 2002--2007, the sensible heat transport is larger than the potential energy transport.
	
	In the year 1988 (Fig.~\ref{fg:energy_physics}a), the total energy is less than in the years 2002--2007.  The 1988 diabatic heating is equinoctial, with the maximum heating near the equator.  Thus, we would expect energy transport away from the equator, i.e., positive in the Northern Hemisphere and negative in the Southern Hemisphere.  Indeed, this behavior occurs in the potential energy term.  However, the sensible heat transport opposes this motion and is large enough to make the total energy positive everywhere.  Since the equinox for other planetary atmospheres such as Earth or Mars is a transition period, it may be that many non-zero transport processes are occurring to nearly cancel each other out.  A second explanation is that in the PGCM, there is an additional energy source when the temperature falls below the freezing temperature.  The temperature is instantaneously snapped back to the freezing temperature, implying the addition of latent heat in the atmosphere.  We suspect that in the 1988 model results, the latent heat term is so strong that it is dominating the other terms, and speculate that when an explicit volatile cycle is included, it will be important near equinox.
		
\begin{figure}
\begin{center}
 \noindent\includegraphics[width=1.0\textwidth]{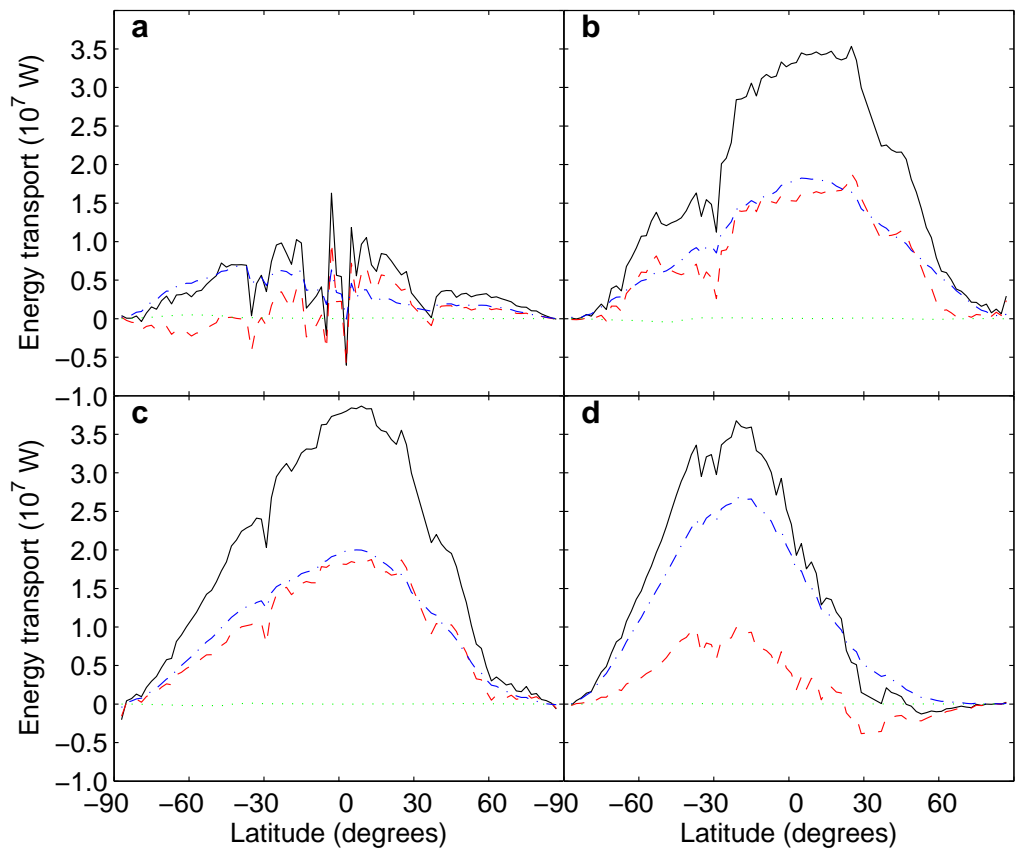}

 \caption{\label{fg:energy_comp}  Simulated vertically averaged meridional transport of energy on Pluto corresponding to (a) 9 June 1988 (b) 21 August 2002 (c) 12 June 2006 (d) 31 July 2007.  The black solid line is the total energy transport, the red dashed line is the mean meridional transport, the green dotted line is the stationary eddy transport, and the blue dashed-dotted line is the transient eddy transport.  These terms are defined in Eq.~\ref{eq:eterms} and have been multiplied by a factor of $2\pi a \cos\phi p_s$ (where $a$ is the surface radius, $\phi$ is latitude, and $p_s$ is surface pressure) to obtain units of Watts.  The results have been time averaged over 90 Earth days.  Positive values indicate northward transport of heat.}

 \end{center}
  \end{figure}

	For $v$ and some variable $A$ (i.e., $T$, $z$ and $(u^2+v^2)$ in Eq.~\ref{eq:energy}), we may use the relationship
	\begin{equation}\label{eq:eterms}
	\left[\overline{vA}\right]=\left[\overline{v}\right]\left[\overline{A}\right]+\left[\overline{v}^*\overline{A}^*\right]+\left[\overline{v'A'}\right],
	\end{equation}
	where stars denote a departure from the zonal mean and primes denote a departure from the temporal mean.  From left to right, the terms on the right hand side of the equation are the transport associated with the mean meridional circulation, stationary eddies, and transient eddies.  Figure~\ref{fg:energy_comp} shows the breakdown of the total energy into the terms of Eq.~\ref{eq:eterms}.  In all four years, the the contribution from stationary eddies is negligible, which is not unexpected since we have included no longitudinal asymmetries in the topography, surface albedo, emissivity, or surface thermal inertia fields.  In 2002 and 2006 the mean meridional circulation and transient eddies are of comparable magnitude, while in 2007 the transient term is larger.  Again in 1988 there is no clear pattern for the energy transport terms.

	\subsection{Triton GCM results}
\begin{figure}
\begin{center}
 \noindent\includegraphics[width=0.7\textwidth]{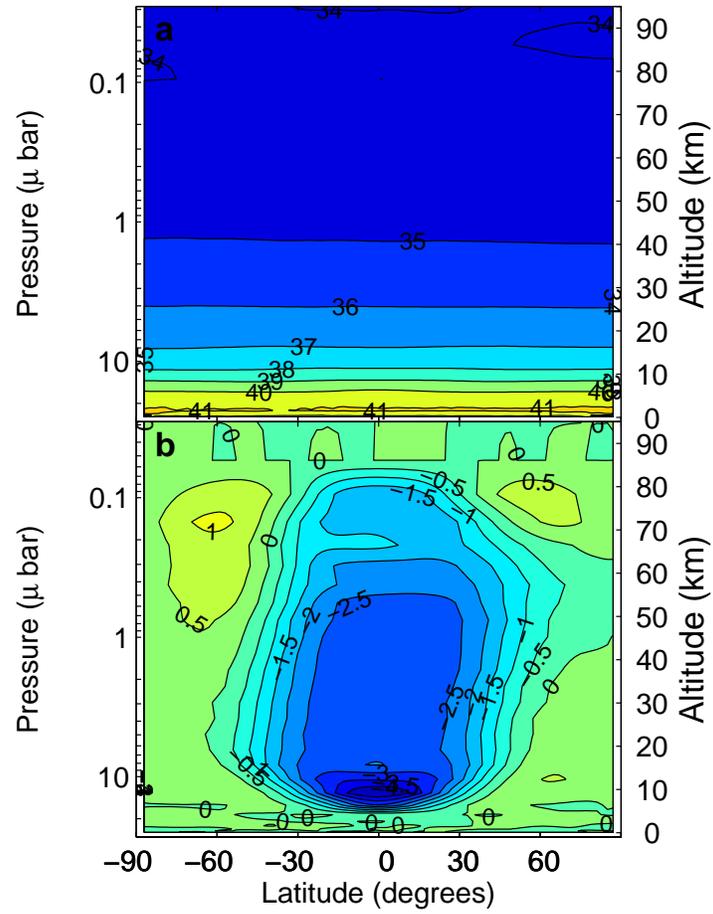}

 \caption{\label{fg:trgcm}  TrGCM results from 4 November 1997.  (a) Zonally averaged temperature (K) and (b) zonally averaged zonal wind (m~s$^{-1}$).}

 \end{center}
  \end{figure}

	Figure~\ref{fg:trgcm} shows the zonally averaged temperature and zonal wind TrGCM results for 4 November 1997.  Triton's globally averaged temperature is much colder (over 30~K) than Pluto and the temperature decreases with height rather than increases, due to the assumption that the \methane~mixing ratio is lower by a factor of 50.  Note that other radiative-conductive models of Triton~\citep[e.g.][]{elliot:2000} exhibit temperature increasing with height, as does ours when a higher \methane~mixing ratio is used. In the TrGCM assumptions, we have incorporated more recent occultation data from~\citet{elliot:1997} and spectral data from \citet{lellouch:2010}, which results in a lower \methane~mixing ratio.  On the other hand, inversion temperatures derived by \citet{elliot:2003b} of the 1997 HST light curve do not show an inversion in the region from 17 to 57~km altitude.  Instead, the temperature profiles show a complicated vertical structure, but lie in a vary narrow temperature range (51--53~K).  However, this temperature is still warmer than our results.

	 Like in the PGCM results, the meridional temperature gradient in our TrGCM results is weak~\citep[as in][]{yelle:1995}.  The zonally averaged zonal wind contains an easterly jet in the tropics and midlatitudes that extends from the top of the assumed troposphere (8~km) to the bottom of the Rayleigh friction layer (near 86~km altitude).  The jet has a maximum magnitude of 4~\ms~at the equator and around 10~km altitude (10~$\mu$bar) pressure.  The zonal winds at the poles are weakly westerly and achieve a maximum magnitude at 70--80~km.  Again, the winds in the frictional boundary layer, which happens to also extend to 8~km, are weak.
	 
	 Voyager 2 detected two well-observed plumes~\citep{smith:1989}, denoted as the ``west'' and ``east'' plumes, located at $-50$\de~and $-57$\de~latitude, respectively.  The west plume appears as a dark column that abruptly terminates at 8~km altitude.  Connected to this plume is a more diffuse clouds that extends westward for at least 150~km.  Similarly, the east plume also rises to an altitude of 8~km and is directed westward at altitude.  These plumes act as a tracer and indicate that the flow is easterly, which agrees with the TrGCM results.  \citet{hansen:1990} identified a terminator cloud in Voyager 2 images that moved 13\ms eastward and was located at an altitude of 5~km.  The direction and speed of this cloud is opposite in direction and much higher in speed than indicated by the TrGCM results.  However, \citet{hansen:1990} point out that because the inferred velocity of this cloud is so similar to the velocity of the terminator of Triton's surface, the cloud may in fact be a stationary, elongated east-west cloud that is being illuminated at different points along the cloud by the low altitude sunlight.  This interpretation would be consistent with the TrGCM results, since in the Southern Hemisphere (the observed location of the cloud), the winds are nearly stationary.
	 
\citet{hansen:1990} also observed several crescent streaks at 1--3~km altitude and surface ($<1$~km altitude) streaks that potentially indicate wind direction at these altitudes.  The direction of the crescent streaks was nearly uniformly westward, while the surface streaks were variable but mainly northeastward.  In the lowest layer of the TrGCM, there is no favored direction at the latitudes and longitudes of the crescent and surface streaks; small- to medium-scale structures of convergence/divergence and rotation are present.  Note that the TrGCM is designed to model the full depth of the atmosphere and has a surface drag scheme that represents the bulk effect of the frictional boundary layer on the free atmosphere.  To explain the crescent and surface streaks seen by Voyager 2, a mesoscale model the better represents small-scale turbulence near the surface is required.  Thus, it is not appropriate to compare the crescent and surface streaks with the TrGCM results. 
	 
	 Like in the PGCM results, in the TrGCM results there is practically no vertical motion and the surface pressure variation is small compared to the globally averaged value~\citep[in agreement with][]{yelle:1995}.  Note that the vertical rising motion observed within Triton's surface plumes is not inconsistent with our result for a number of reasons.   First, if the plume material is composed of a different material and/or is warmer than the background atmosphere, this would allow the plumes to be buoyant.  Second, the plumes may move vertically if they have a non-zero upward initial velocity~\citep[see][for a detailed discussion of plume sources]{kirk:1995}.  Thirdly, our GCM resolution is much coarser than the horizontal area of the plumes.  The GCM may be averaging out any upward motion from the plumes with downward motion elsewhere in the grid box, resulting in zero net velocity.
	
\begin{figure}
\begin{center}
 \noindent\includegraphics[width=0.7\textwidth]{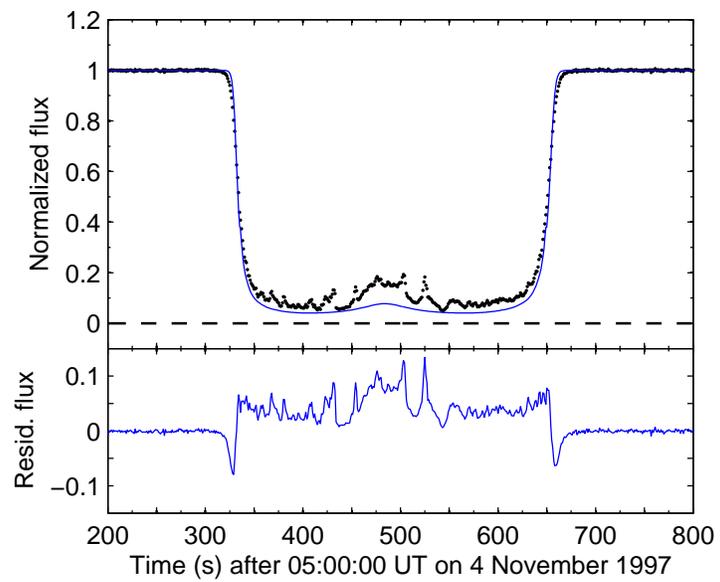}

 \caption{\label{fg:lc1997}  Top panel: light curve corresponding to the TrGCM (blue curve) and HST data (red points) on 4 November 1997.  The dashed horizontal line is the zero flux level.  Bottom panel: residual between data and model. }

 \end{center}
  \end{figure}

	Figure~\ref{fg:lc1997} shows the calculated TrGCM light curve calculated and the light curve data from the HST on 4 November 1997.  The model systematically underestimates the data near the midtime by 5--10$\%$, unlike the Pluto model light curves where the opposite was the case.  Again, the lower atmosphere is apparently not being simulated accurately.  An underestimate by the model cannot be accounted for by haze, but could be the result of the vertical temperature gradient dropping off too strongly with altitude (i.e., the true lower atmospheric vertical temperature gradient is more isothermal or contains an inversion).  We have investigated the model underestimate by performing runs with surface pressure doubled, \methane~mixing ratio increased by an order of magnitude, tropopause height doubled, and surface temperature fixed at 52~K~\citep[following][]{elliot:2003b}.  None of these modifications were successful at increasing the model light curve flux near midtime.
	
	\section{Discussion}\label{se:discussion}
	\subsection{The effect of Pluto's lapse rate on circulation}
	In the 2D and 3D PGCMs (also the TrGCM), we have found practically no vertical motion.  This is unique versus other planetary atmospheres in the Solar System, such as Earth, Venus, Mars, Titan, and the giant planets.  These planets have one or more overturning circulation patterns such as Hadley, Walker, or eddy driven circulations.  It is a well-known fact that the atmosphere resists vertical motions when a statically stable (subadiabatic) vertical temperature gradient is present.  In this section, we show that the temperature inversion in Pluto's stratosphere appears to be what is preventing these vertical motions.
	
	To demonstrate the effect of a less statically stable atmosphere on Pluto's atmospheric circulation, we replace the non-LTE, shortwave dominated \citet{yelle:1989} radiative-conductive forcing with a simple radiative-convective scheme that is representative of a greenhouse dominated atmosphere such as Mars or Earth.  The external heating term is given by
	\begin{equation}
	\frac{\partial T}{\partial t} = -k_T\left(T-T_{eq}\right),
	\end{equation}
	where $k_T$ is the radiative heating rate and $T_{eq}$ is the equilibrium temperature.  $T_{eq}$ is defined by (Zalucha et al., 2010)
	\begin{equation}\label{eq:trc}
\sigma T_{eq}^4=\begin{cases}
Q_o\left[0.5+0.75 \tau(p)\right] & \tau < \tau_t \\
Q_o\left[0.5+0.75 \tau_t\right]\left[\tau\left(p\right)/\tau_t\right]^{4R/c_p} & \tau \geq \tau_t,
\end{cases}
\end{equation}
	where $Q_o$ is the insolation, $p$ is pressure, $\tau$ is the long wave optical depth, $\tau_t$ is the value of $\tau$ at the top of the convective layer (found by energy balance constraints), and $R$ is the specific gas constant.  Equation~\ref{eq:trc} states that the equilibrium temperature follows an adiabat in the surface convective layer of the atmosphere and a radiative equilibrium profile above.  The long wave optical depth is linearly dependent on pressure as $\tau=\tau_o(p/p_o$), where $\tau_o$ and $p_o$ are reference optical depth and pressure, respectively.
	
	In this example~\citep[following][]{zalucha:2010b}, we assume that the surface pressure (set equal to $p_o$) is equal to 13.2~$\mu$bar, which is an intermediate value for the period 1988--2007.  $\tau_o=2$ and $k_T$ are tuned until Pluto-like temperatures are achieved.  The exact values for $p_o$, $\tau_o$, and $k_T$ are not important; our main goal is simply to drastically alter the temperature stratification.
	
\begin{figure}
\begin{center}
 \noindent\includegraphics[width=0.8\textwidth]{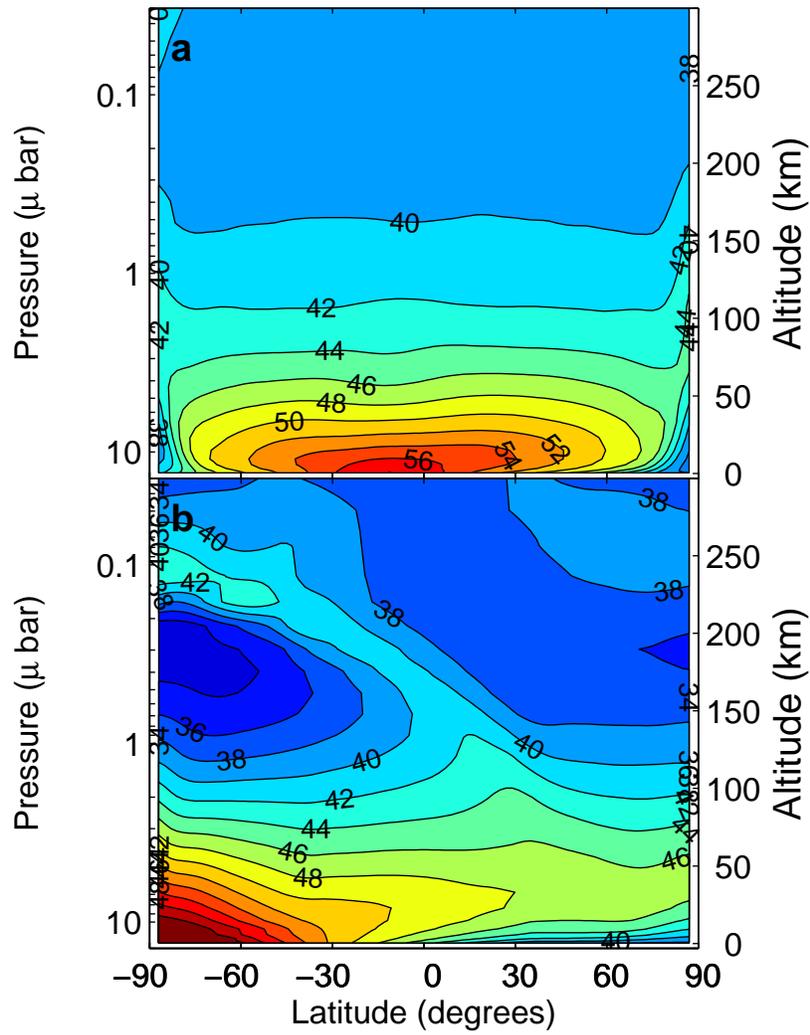}

 \caption{\label{fg:temp_mars}  Zonally and time averaged temperature (K) for the PGCM with greenhouse gas absorber (a) equinox-type case (b) solstice-type case.   The surface pressure for this example is 13.2~$\mu$bar.}

 \end{center}
  \end{figure}

\begin{figure}
\begin{center}
 \noindent\includegraphics[width=0.8\textwidth]{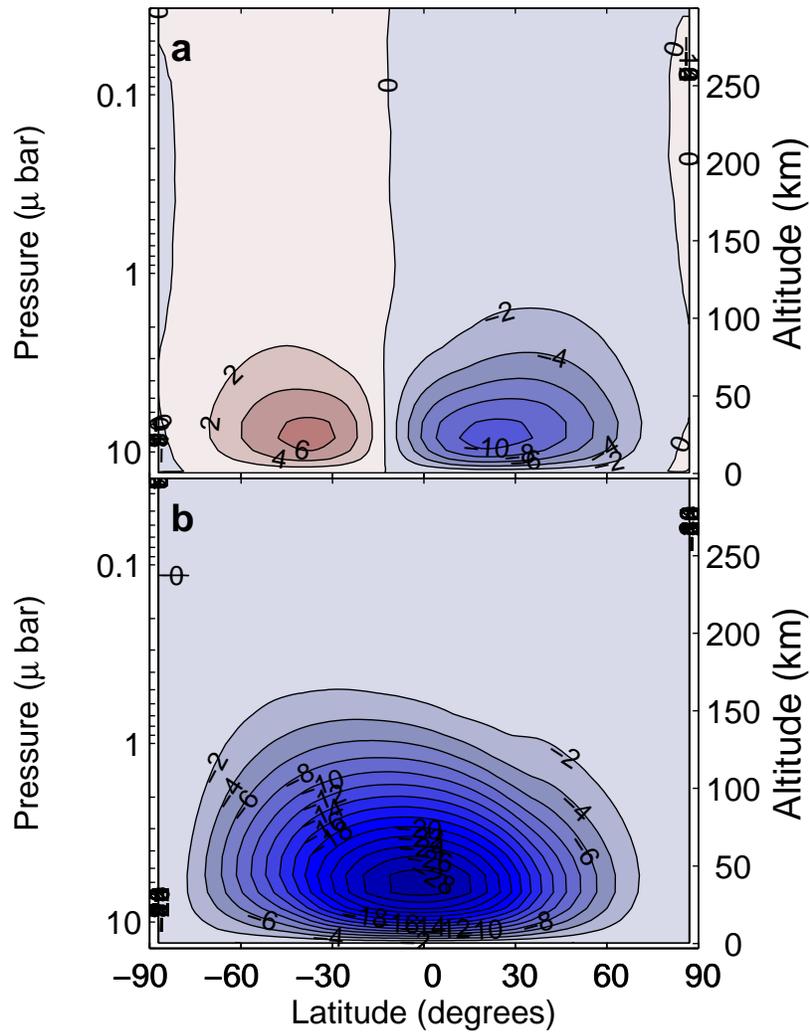}

 \caption{\label{fg:psi_mars}  Zonally and time averaged mass stream function ($10^7$~kg~s$^{-1}$) for the PGCM with greenhouse gas absorber (a) equinox-type case (b) solstice-type case.   Positive flow is counter-clockwise.  The surface pressure for this example is 13.2~$\mu$bar.}

 \end{center}
  \end{figure}

	Figure~\ref{fg:temp_mars} shows a zonally and time averaged latitude-height temperature cross-section for Pluto with a greenhouse dominated atmosphere near equinox and solstice.  The heating profile produces a temperature structure that decreases monotonically with height, although some regions with temperature inversions exist due to dynamical processes.  Figure~\ref{fg:psi_mars} shows the corresponding zonally and time averaged latitude-height cross section of mass stream function.  There is clearly an overturning circulation pattern (Hadley cells);  at equinox there are two cells  (one cell in each hemisphere) and at solstice there is a single planet-wide cell.  It is not surprising that with a Mars-like temperature structure we obtain a Mars-like circulation.  However, the preceding example shows the importance of the vertical lapse rate on atmospheric circulation.  In the nominal PGCM with a steep temperature inversion, the mass stream function is effectively zero everywhere.  It is absolutely critical to include the steep temperature inversion in atmospheric models of Pluto to ensure a circulation with no vertical motions.
	
	In models that include explicit surface-atmosphere mass exchange, a local deficit of mass (i.e., low pressure) is created where surface ice deposition occurs, and a local abundance of mass (i.e., high pressure) is created when sublimation occurs.  On Mars, these pressure gradients drive global flows from high to low pressure at all altitudes.  So-called condensation flows probably exist on Pluto, but they will likely be confined to the atmosphere very near the surface.  The surface exchanges mass with air immediately above the surface, and with the radiative-conductively forced vertical temperature gradient in place, mass will not be able to move upward or downward to or from layers aloft at large scales.  Moreover, the weak dayside-nightside temperature contrast in the PGCM results (due to the slow radiative-conductive relaxation rate compared to the rotation rate) suggests the nightside of Pluto will have a lower frost deposition rate than if there were no atmospheric heat transport.
	
An exception to this scenario is the strength of turbulence and vertical mixing in the atmosphere near the surface.  If the vertical mixing was able to mix sublimating surface volatiles high enough into the atmosphere, then the condensation flow need not be confined to the surface.  Our GCM does not include a parameterization for small-scale turbulence; however its properties on Pluto are not likely to known in the near future anyway since it requires \textit{in situ} measurements and empirical modeling to characterize it, which has currently only been done for Earth and Mars.
	
	The PGCM light curves generated from the simulation with fixed surface temperature at the freezing point had slightly reduced $\chi^2$ values compared with an involatile surface allowed to vary in time and space.  A fixed surface temperature at the freezing point suggests that Pluto has a large (i.e., thick) surface volatile ice reservoir that is able to maintain the freezing temperature despite solar and atmospheric forcing.
	
	\subsection{Triton's lower atmosphere}
	The TrGCM shows that the lowest $\sim$5~km of Triton's equatorial atmosphere is characterized by relatively weak zonal winds, before a sharp increase in wind speed above this altitude.   This behavior is very similar to the Voyager 2 observation of plumes on Triton that extended to 8~km altitude before being blown downstream~\citep{smith:1989}.  \citet{smith:1989} suggested that the behavior of the plumes is controlled by a temperature inversion at the tropopause and/or vertical structure in the wind speeds. 
	
	  Two important model elements are located in the lowest region of the atmosphere.  First, the radiative-conductive-convective heating/cooling scheme is set so that in the steady state, the atmosphere below the tropopause (fixed at 8~km) follows a moist (with respect to N$_2$) adiabat (i.e., decreases with height).  Second, a frictional layer is present from the surface to a pressure level equal to 70$\%$ of the surface pressure (in this case 17.64~$\mu$bar).  When full model dynamics and radiative-conductive-convective heating/cooling are allowed, we find that the temperature actually increases slightly with height in the lower atmosphere.  To test whether it is the external heating/cooling or surface friction that is dominating the lower atmosphere, we performed a simulation where the diffusion of potential temperature was turned off in the radiative-conductive-convective heating/cooling scheme.  This action removes the thermally-induced troposphere.  In this case the near-surface winds were still weak.  Thus, the behavior of the plumes once they have been released into the atmosphere can be caused by strong wind shear capping the lower atmosphere, rather than thermal effects.  This result corresponds to the second intepretation of the plume behavior suggested by \citet{smith:1989}.
	  
	  At the time of the Voyager 2 encounter, \citet{ingersoll:1990} used Ekman layer theory (near-surface balance between the pressure gradient force, Coriolis force, and turbulent drag) to predict an anticyclone at the south pole.  Anticyclonic flow corresponds to flow opposite to the body's rotation, or in this case westerly.  Our TrGCM results do show westerly flow at the south pole (poleward of $\sim$60\de), but the wind speed is less ($< 0.5$~\ms) than the \citet{ingersoll:1990} estimate of 5 to 15~\ms.  Dust devils, which are not resolvable in our model, have also been proposed as another dynamical mechanism to explain the plumes~\citep{ingersoll:1990b}.  However, the static stability of the lower atmosphere of our model would prohibit dust devils, which can only form in the presence of neutral or unstable vertical temperature profiles.

				\section{Conclusion}
			We have simulated the atmospheres of Pluto and Triton using a 3D GCM based on the MIT GCM dynamical core and the \citet{yelle:1989} and Zalucha et al. (2011b) radiative-conductive-convective schemes.  We find that, like the 2D version of this GCM, there is no significant large-scale vertical motion and hence no overturning cells in the Pluto or Triton atmospheres.  Triton's plumes, which are not explicitly present in our model, can be an exception to this rule if they are composed of a different and/or more buoyant material than the background atmosphere or escape the surface with a non-zero vertical velocity.
			
			Some meridional motion is present in our simulations (maximum magnitude 2~\ms) in the form of transient waves and mean circulations that transport heat meridionally very efficiently.  As a result, the meridional temperature gradient is weak (as is the dayside-nightside temperature contrast).  On both Pluto and Triton, a zonal jet is located at the equator and midlatitudes.  Pluto exhibits a more complicated vertical structure in the zonal winds with three distinct regimes present.
			
			The strong temperature inversion in Pluto's lower atmosphere prevents large-scale overturning cells.  By substantially changing the lapse rate so that it was similar in vertical structure to a greenhouse-dominated atmosphere (e.g., Mars), we were able to recover Hadley cells.  We also showed that the macroscopic behavior of plumes on Triton can be explained by the dynamical properties (i.e., strong wind shear) of the lower atmosphere alone, since a layer of abruptly strong winds overlayed a surface layer of essentially quiescent winds when the convective component of the heating external scheme was removed from the TrGCM.
			
			The PGCM and TrGCM are continually improving.  In the future we plan to include \methane~transport, a multi-layer surface model, and a frost cycle.  The latter two items represent particular challenges because the surface and subsurface parameters (e.g., albedo, surface thermal inertia, and emissivity) represent a particularly large parameter space that is not well constrained~\citep{hansen:1996,young:2012}.  These may also require multi-year simulations to equilibrate the surface frost thickness and/or subsurface temperature (heat storage).
			
			Because Pluto and Triton are so similar in their bulk properties it is useful to study them in tandem.  A relatively large amount of data has been obtained through observations of stellar occultations by Pluto, and more are planned to be acquired by the New Horizons spacecraft, currently en route to Pluto.  While Triton has already been visited by Voyager 2, those observations are now over two decades old and Triton's atmosphere has potentially undergone seasonal changes.  With further observations of Triton, such as stellar occultations, spectra, and surface albedo, we will be able to better constrain the TrGCM, and have the secondary benefit of improving the PGCM.

%% Using an acknowledgements command is not in the Elsevier template,
%% but it can be used.
%\ack
%This work has made use of NASA's Astrophysics Data System.  It 
%also benefitted tremendously from \citet{latexguide}.

%\ack doesn't work.  
\section*{Acknowledgments}
This work used the Extreme Science and Engineering Discovery Environment (XSEDE), which is supported by National Science Foundation grant number OCI-1053575.  We thank two anonymous reviewers who provided suggested revisions to the original manuscript.

\nocite{zalucha:2012a}
\nocite{zalucha:2013}
\nocite{zalucha:2011a}
\nocite{zalucha:2010}
\nocite{zalucha:2011b}

%shouldn't this at least go after the references?  Or does ``number of manuscript pages'' refer to the entire paper?
%\label{lastpage}

%\appendix

%don't use the built in appendix environment, because we only have 1 appendix 
%\renewcommand{\theequation}{A-\arabic{equation}}
%  % redefine the command that creates the equation no.
%  \setcounter{equation}{0}  % reset counter 
%  \section*{Appendix: An implicit scheme for a general one-dimensional dimensional diffusive equation with varying step size}

% Bibliographic references with the natbib package:
% Parenthetical: \citep{Bai92} produces (Bailyn 1992).
% Textual: \citet{Bai95} produces Bailyn et al. (1995).
% An affix and part of a reference:
%   \citep[e.g.][Ch. 2]{Bar76}
%   produces (e.g. Barnes et al. 1976, Ch. 2).-

\label{lastpage}

\clearpage	% Make sure things don't run together.

\begin{table}[h!]
\begin{center}

 \caption{
	\label{tb:params}	
	\textbf{Pluto and Triton GCM parameters }
	}
\begin{tabular}{lll}
\hline 
\hline
 Parameter & Pluto value & Triton value\\
 \hline
 Surface radius (km) & 1152$^{a}$ & 1353\\
 Orbital eccentricity & 0.251 & 0.0097$^b$ \\
 Semimajor axis (AU) & 39.8 & 29.9$^b$ \\
 Rotation rate ($10^{-5}$~s$^{-1}$) & $-1.13856$ & $-1.2374$\\
 Obliquity with respect to Sun (\de) & 60.4 & 28.3\\
 Surface gravitational acceleration (m~s$^{-1}$) & 0.63 & 0.77 \\
 Wavelength-integrated solar constant (W~m$^{-2}$) & 0.864 & 1.137\\
 Ecliptic longitude of perihelion (\de) & 186 & 38.0 \\
\hline
\multicolumn{3}{l}{$^a$different from ZG12.} \\
 \multicolumn{3}{l}{$^b$of Neptune with respect to Sun.}
\end{tabular}
\end{center}
\end{table}

\begin{table}[h!]
\begin{center}

 \caption{
	\label{tb:results_free}	
	\textbf{Best-fit PGCM parameters (surface temperature free to evolve)}
	}
\begin{tabular}{lcccc}
\hline 
\hline
 Event & Global mean surface & \methane~mixing  & Minimum        & Degrees of      \\
       &  pressure ($\mu$bar) & ratio ($10^{-3}$) & reduced $\chi^2$ & freedom  \\
 \hline
 1988 KAO & 12$\pm$4 & 6$\pm$4 & 6.863  & 998 \\ 
 2002 UH 2.2m & 24$\pm$4 & $\ge 10$ & 2.276 & 2390 \\
 2006 AAT & 24$\pm$4 & $\ge 10$ & 6.773 & 3598  \\
 2006 Siding Spring & 24$\pm$24 & $\ge 10$ & 1.438 & 1798 \\
 2007 Mt. John & $\ge 28$ &  6$\pm$4 & 4.343 & 1999 \\
 \hline
\end{tabular}
\end{center}
\end{table}

\begin{table}[h!]
\begin{center}

 \caption{
	\label{tb:results_fixed}	
	\textbf{Best-fit PGCM parameters (surface temperature fixed)}
	}
\begin{tabular}{lcccc}
\hline 
\hline
 Event & Global mean surface & \methane~mixing  & Minimum     & Degrees of        \\
       &  pressure ($\mu$bar) & ratio ($10^{-3}$) & reduced $\chi^2$ & freedom \\
 \hline
 1988 KAO & 8 $\pm$ 4 & $\ge 10$  & 5.336 & 998 \\ 
 2002 UH 2.2m & 20$\pm$ 4 & $\ge 10$ & 1.994 & 2390\\
 2006 AAT  & 20$\pm$ 4 & $\ge 10$  & 5.259 & 3598 \\
 2006 Siding Spring & 20$\pm$ 4 & $\ge 10$ & 1.338 & 1798 \\
 2007 Mt. John  & 16$\pm$ 4 & $\ge 10$ & 2.210 & 1999\\
 \hline
\end{tabular}
\end{center}
\end{table}

\begin{table}[h!]
\begin{center}

 \caption{
	\label{tb:2007example}	
	\textbf{Selected reduced $\chi^2$ for the 2007 fixed surface temperature case}
	}
\begin{tabular}{ccc}
\hline 
\hline
    Global mean surface  &Reduced $\chi^2$ for & Reduced $\chi^2$ \\
 pressure ($\mu$bar)     & \methane~mixing  & \methane~mixing       \\
                         & ratio of 0.6$\%$ & ratio of 1$\%$ \\
 \hline
 12 										 & 2.386         &   2.217 \\
 16 										 & 2.332          &   2.210 \\
 20 										 & 2.332          &   2.255 \\
 \hline
\end{tabular}
\end{center}
\end{table}

\begin{table}[h!]
\begin{center}
\caption{
	\label{tb:chi2vsrs}	
	\textbf{Minimum reduced $\chi^2$ for selected surface radii (surface temperature fixed)}
	}
\begin{tabular}{lcccc}
\hline 
\hline
 Event & 1130~km & 1152~km  & 1180~km     & Degrees of freedom        \\
 \hline
 1988 KAO & 6.492 & 5.336 & 5.367 & 998\\
 2002 UH 2.2m & 2.000 & 1.994 & 2.379 & 2390\\
 2006 AAT & 6.046 & 5.259 & 7.142 & 3598\\
 2006 Siding Spring & 1.391 & 1.338 & 1.648 & 1798   \\ 
 2007 Mt. John & 2.269 & 2.210 & 2.384 & 1999\\
 \hline
\end{tabular}
\end{center}
\end{table}

\clearpage

%% --Figures-- %%

% figure 1

  %figure 2

%figure 3

 % figure 4

  %figure 5
  
  %figure 6
  
  %figure 7
  
  %figure 8
  
  %figure 9
  
  %figure 10
  
  %figure 11
  
  %figure 12

  %figure 13
  
%why why why doesn't this work
%\label{lastpage}

\end{document}